%% file: main_icde.tex
\documentclass[conference]{IEEEtran}
\IEEEoverridecommandlockouts
\usepackage[normalem]{ulem}
\useunder{\uline}{\ul}{}
\usepackage[ruled,vlined,linesnumbered]{algorithm2e} 
\usepackage{booktabs}
\usepackage{cite}
\usepackage{algorithmic}
\usepackage{graphicx}
\usepackage{textcomp}
\usepackage{xcolor}
\usepackage{graphicx}
\usepackage{subcaption}
\usepackage{amsmath,amssymb,amsfonts}
\usepackage[misc]{ifsym}

\usepackage{amsthm}
\newtheorem{theo}{Theorem}

\newtheorem{corl}{Corollary}
\newtheorem{defi}{Definition}

\def\BibTeX{{\rm B\kern-.05em{\sc i\kern-.025em b}\kern-.08em
    T\kern-.1667em\lower.7ex\hbox{E}\kern-.125emX}}

\begin{document}

\title{PICO: Accelerating All $k$-Core Paradigms on GPU}

\author{\IEEEauthorblockN{Chen Zhao$^{\S}$, Ting Yu$^{\dag}$, Zhigao Zheng$^{\S, \textrm{\Letter}}$, Song Jin$^{\S}$, Jiawei Jiang$^{\S}$, Bo Du$^{\S}$, Dacheng Tao$^{\ddag,\textrm{\Letter}}$}
\IEEEauthorblockA{
 $^{\S}$School of Computer Science, Wuhan University, Wuhan 430072, China\\
 $^{\S}$National Engineering Research Center for Multimedia Software, Wuhan University, Wuhan 430072, China\\
 $^{\S}$Institute of Artificial Intelligence, Wuhan University, Wuhan 430072, China\\
 $^{\S}$Hubei Key Laboratory of Multimedia and Network Communication Engineering, Wuhan University, Wuhan 430072, China\\
 $^{\S}$Hubei Luojia Laboratory, Wuhan 430072, China\\
$^{\dag}$ Zhejiang Lab, Hangzhou, China\\
$^{\ddag}$ UBTECH Sydney Artificial Intelligence Center, University of Sydney, Australia\\
$^{\ddag}$ School of Information Technologies, University of Sydney, Australia\\
\{zhaochen13,zhengzhigao,jinsong,jiawei.jiang,dubo\}@whu.edu.cn, yuting@zhejianglab.com, dacheng.tao@gmail.com}
}

\maketitle

\begin{abstract}
Core decomposition is a well-established graph mining problem with various applications that involves partitioning the graph into hierarchical subgraphs. Solutions to this problem have been developed using both bottom-up and top-down approaches from the perspective of vertex convergence dependency. However, existing algorithms have not effectively harnessed GPU performance to expedite core decomposition, despite the growing need for enhanced performance. Moreover, approaching performance limitations of core decomposition from two different directions within a parallel synchronization structure has not been thoroughly explored.
This paper introduces an efficient GPU acceleration framework, PICO, for the \emph{Peel} and \emph{Index2core} paradigms of $k$-core decomposition. We propose \emph{PeelOne}, a \emph{Peel}-based algorithm designed to simplify the parallel logic and minimize atomic operations by eliminating vertices that are 'under-core'. We also propose an \emph{Index2core}-based algorithm, named \emph{HistoCore}, which addresses the issue of extensive redundant computations across both vertices and edges. Extensive experiments on NVIDIA RTX 3090 GPU show that \emph{PeelOne} outperforms all other \emph{Peel}-based algorithms, and \emph{HistoCore} outperforms all other \emph{Index2core}-based algorithms. Furthermore, \emph{HistoCore} even outperforms \emph{PeelOne} by $1.1\times \sim 3.2\times$ speedup on six datasets, which breaks the stereotype that the \emph{Index2core} paradigm performs much worse than the \emph{Peel} in a shared memory parallel setting.

\end{abstract}


\input{sec1_intro.tex}
\input{sec2_banckground.tex}
\input{sec3_peelone.tex}

\input{sec4_hist.tex}

\input{sec5_exp.tex}
\input{sec6_relatedworks.tex}
\input{sec7_conclusion.tex}

\section*{Acknowledgment}

This work is supported in part by the National Natural Science Foundation of China under Grant 62225113, 62372333, the Fundamental Research Funds for the Central Universities under Grant 2042023kf0135, the Key Research and Development Program of Hubei Province under Grant 2023BAB078, the Project funded by China Postdoctoral Science Foundation under Grant 2022M722459, the Natural Science Foundation of Hubei Province under Grant 2022CFB795, the Knowledge Innovation Program of Wuhan - Basic Research under Grant 2023010201010063, and the Natural Science Foundation of Zhejiang Province under Grant LQ22F020033.

\bibliographystyle{IEEEtran}
\input{main_icde.bbl}

\end{document}

%% file: sec1_intro.tex
\section{INTRODUCTION}

\begin{figure}[ht]
    \centering
    \includegraphics[width=6cm]{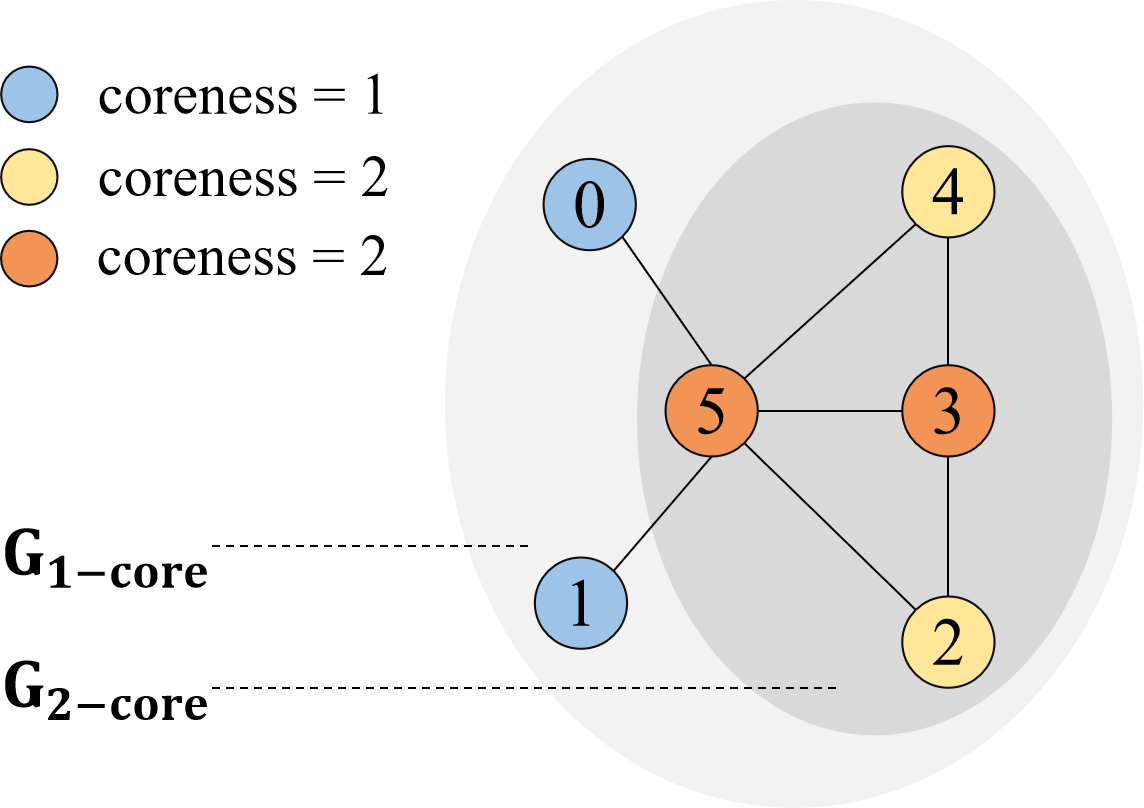}
    \caption{An illustration of $k$-cores and coreness resulted from core decomposition in the example graph $G_1$.}
    \label{fig:k-core}
\end{figure}

Given a graph $G = (V, E)$, for an integer $k$, a $k$-core is a maximum subgraph of $G$ with all the vertices degree $\geq k$. The coreness of vertex $v \in G$ is the maximum value of $k$ for which there is a $k$-core that contains $v$. The target of core decomposition is to determine the coreness of each vertex $v \in G$. We illustrate core decomposition in Fig.~\ref{fig:k-core}. The entire graph is a 1-core, while vertices $\{v_2, v_3, v_4, v_5\}$ form the 2-core. No 3-core is present. Therefore, the coreness of vertices $\{v_0, v_1\}$, and $\{v_2, v_3, v_4, v_5\}$ are 1 and 2, respectively.

Due to the simple and elegant structure with linear complexity~\cite{peel-paradigm}, the $k$-core is widely used in many applications. In social networks, researchers employ hierarchical subgraph processing to accelerate intensive graph clustering~\cite{corecluster}, clique finding~\cite{cliquefinding}, and community detection~\cite{communitydetection1, communitydetection2} and search~\cite{communitysearch}. The coreness can help user engagement, prevent unraveling and improve network stability in social network~\cite{anchored-core1, anchored-core2, anchored-core3, anchored-core4}. Furthermore, $k$-core is an effective tool to predict and visualize the functions of complex structures in biology or ecology~\cite{Protein1, Protein2}. Numerous studies explore core decomposition in diverse networks with rich semantics, such as directed graphs, uncertain graphs, dynamic graphs and others ~\cite{d-core2, uncertain-core4, dyna-core6, h-core1}. 

There are many algorithmic techniques for core decomposition in different settings~\cite{overview-core}, originating from the initial proposal of the $k$-core concept by Seidman~\cite{peel-paradigm}. These techniques can be classified into two paradigms: \emph{Peel}~\cite{peel-paradigm} and \emph{Index2core}~\cite{index2core-paradigm}. In the \emph{Peel} paradigm, the algorithm iteratively removes vertices with the minimum degree until all the coreness values are obtained. In the \emph{Index2core} paradigm, the $h$-index value of each vertex is computed iteratively until convergence to the coreness is achieved. From the perspective of vertex convergence dependency, the \emph{Peel} paradigm is bottom-up, while the \emph{Index2core} paradigm is top-down.

With the continuous growth of the data scale and the widespread application of $k$-core, the pursuit of optimal performance remains ongoing. Graphics Processing Units (GPUs) have gained significant popularity for accelerating graph processing algorithms and applications due to their excellent parallel computing capabilities and memory bandwidth.
To explore core decomposition in a massively parallel setting, we have conducted research on various works that focus on accelerating the $k$-core algorithm on GPU~\cite{GPU-core1, GPU-core3, GPU-core4, gunrock}. 

Though significant efforts have been made to improve the efficiency of the \emph{Peel} paradigm on GPU, the existing works are implemented at different programming levels, including the latest work~\cite{GPU-core4}.
Therefore, they do not reveal the key optimal parallelization design of the core logic of the \emph{Peel} algorithm. One observation is that the complete peeling process on the objective graph is overly redundant for calculating the coreness of vertices. For a vertex with a coreness value of $k$, the coreness can be identified when its degree is reduced to $k$. However, the peeling process aims to reduce every edge in the objective graph. During the reducing, the residual degree of some vertices might be different from the coreness value. This leads to some unnecessary atomic computational overhead and storage costs.

On the other hand, the parallelism potential of \emph{Index2core} on GPU has not been thoroughly explored.
\emph{Index2core} gradually approximates the coreness value of each vertex by iteratively estimating its $h$-index.
In each iteration, the $h$-index value change of a frontier is estimated based on the latest values of its neighbors from the previous iteration. 
Thus, existing \emph{Index2core} methods regard its neighbors as new frontiers in the next iteration, once the $h$-index of current frontier is changed.
However, only a small portion of neighbors will actually be affected by the changes in the $h$-index of frontiers, resulting in a large number of mistaken frontiers (the estimation remains unchanged) in the next iterations.
Moreover, the computational cost of the $h$-index update operation for a frontier is positively correlated with the number of its edge accessing (i.e., degree), so that high-degree vertices incur heavier computational costs.
Those high-degree vertices, potentially requiring more iterations to converge to their coreness, become `multi-changed' vertices. Consequently, a small number of multi-changed vertices end up with a significantly larger computational workload, leading to an imbalance in the workload distribution and becoming a bottleneck for GPU parallel computing.

In real scenarios, the \emph{Peel} paradigm is highly sensitive to the dependency order among nodes and is suitable for in-memory computation on static graphs, while the \emph{Index2core} paradigm is suitable for computations on dynamic graphs and graph partition issues. Therefore, optimizing both paradigms on GPU holds significant importance. The \emph{Peel} methods obtain the coreness of each vertex through a simple judgment of its residual degree, while the \emph{Index2core} methods determine the coreness of each vertex via complex updates, which involve accesses and calculations of its neighbors, once or more. Therefore, the total computational overhead required by the \emph{Peel} methods is lower, closer to the theoretical upper limit of computational complexity for core decomposition problems, and is easier to adapt for effective parallel computation using GPU. Improving the performance of \emph{Index2core} on GPU to levels comparable to those of the optimized \emph{Peel} method serves as an important indicator for assessing the optimization of the Index2core algorithm.

To address the issues mentioned above, we develop a synchronous computing framework that integrates all the optimizations for core decomposition on GPU. This framework supports both the \emph{Peel} and \emph{Index2core} paradigms.
We consolidate various existing optimization methods to abstract the optimal parallel \emph{Peel} paradigm. Within this paradigm, we define the under-core vertex and analyze its impact on the performance. We devise a low-cost elimination method to reduce the redundant atomic operations on under-core vertices. We utilize a dynamic queue to manage frontiers within the same core hierarchy, thereby minimizing the number of iterations required.
For the \emph{Index2core} paradigm, we reduce the number of vertices involved in computing $h$-index by locating the frontiers. By decoupling the construction of the histogram array from the computation of the $h$-index for the multi-changed frontiers, we effectively reduce the number of redundant edge accesses. 

In summary, the contributions of this paper are as follows.

\begin{itemize}
    \item We propose a framework, called PICO, for all k-core paradigms on GPU, which includes the optimal \emph{Peel}-based algorithm \emph{PeelOne} and the \emph{Index2core}-based algorithm \emph{HistoCore} and incorporates the key optimization techniques.
    
    \item In the bottom-up pattern, \emph{PeelOne} simplifies the parallel logic of peeling the objective graph and reduces the atomic operation by the proposed \emph{assertion} method to eliminate the under-core vertex.
    
    \item In the top-down pattern, \emph{HistoCore} locates the frontiers precisely and reduces redundant memory access on the edges of multi-changed frontiers by using the up-to-date histogram information.

    \item Experimental results demonstrate that in their respective paradigms, \emph{PeelOne} and \emph{HistoCore}, as facilitated by the framework, achieve optimal performance across 24 datasets. Furthermore, \emph{HistoCore} outperforms \emph{PeelOne} within the shared-memory parallel environment on six datasets, highlighting the great parallel potential of the \emph{Index2core} paradigm as same with the \emph{Peel} paradigm.
\end{itemize}

The remaining sections are organized as follows.
Section~\ref{background} introduces the background of the two $k$-core decomposition paradigms and the motivation of this paper, which is supported by a set of experiments on GPU. The detailed techniques of the proposed \emph{PeelOne} and \emph{HistoCore} algorithms are presented in Section~\ref{peelone} and Section~\ref{hist}, respectively. Section~\ref{exp} evaluates the performance of the proposed algorithms, and Section~\ref{relatedworks} introduces the existing relevant work. Finally, the paper concludes in Section~\ref{sec:conclusion}.

%% file: sec2_banckground.tex
\section{Background and Motivation}
\label{background}
In this section, we first analyze the features of the two widely used paradigms of the core decomposition algorithm. Then, we introduce the key concepts of graph algorithms on GPU. At last, by revisiting the existing $k$-core on GPU, we conduct a set of experiments on GPU to identify the key factors that limit the parallel efficiency of the two paradigms, which motivate us to design the new framework PICO to accelerate the core decomposition.

\subsection{The $k$-core Decomposition Paradigms}
The objective of $k$-core decomposition is to calculate the coreness value for all vertices in $V$. Table~\ref{tab:symbols} lists the symbols frequently used in this paper. 

\begin{table}[b]
    \caption{Frequently used symbols.}
    \label{tab:symbols}
    \centering
    \small
    \begin{tabular}{lp{6.2cm}}
        \toprule
        Symbols       & Explanations \\
        \midrule
        $G = (V, E)$  & The object graph. \\
        $v_i$         & The vertex with id $i$.\\
        $nbr(u, G)$   & The neighbors of vertex $u$ in $G$.\\
        $deg(u, G)$   & The degree of vertex $u$ in $G$.\\
        $V_k(G)$      & The vertex set in the $k$-core of $G$, $V_k$ for short.\\
        $G_k(V_k)$    & The induced subgraph of $V_k$ in $G$, $G_k$ for short.\\
        $deg_r(u, k, G)$ & The residual degree of vertex $u$ when locating the $k$-core.\\ 
        $V_{res}$  & The residual vertices of $G$ in \emph{Peel} paradigm.\\
        $G_{res}$  & The residual graph of $G$ in \emph{Peel} paradigm.\\
        $core(u, G)$  & The coreness of vertex $u$ in $G$, $core(u)$ for short.\\
        $h^t_u$       & The $h$-index of vertex $u$ in the $t$-th iteration. \\
        $cnt(u, t)$   & The number of neighbors with an $h$-index value no smaller than vertex $u$ in the $t-1$-th iteration in \emph{Index2core} paradigm.\\
        \bottomrule
    \end{tabular}
\end{table}

Most of the existing $k$-core decomposition algorithms can be classified into two paradigms: \emph{Peel} and \emph{Index2core}. In the \emph{Peel} paradigm, the vertices with degree $\leq k-1$, along with their corresponding edges, are removed in each iteration until the $k$-core is located. Initially, vertices with degree 1 are removed and their coreness is set to 1. This process is iterated, increasing the value of $k$, until no vertices remain in the graph. In practice, a flag is used to mark the removed vertices and the degree of their neighbors is reduced to obtain the residual graph. The detailed execution procedure is shown in Algorithm~\ref{Core-Number-peel}. For the \emph{Index2core} paradigm, every vertex estimates its coreness from the latest coreness estimation of all its neighbors by the computation of $h$-index until the estimation of every vertex is unchanged. For a given vertex $v$, the $h$-index is defined as the highest value of $h$ for which $v$ has at least $h$ neighbors with a degree $\geq h$. The detailed execution procedure is shown in Algorithm~\ref{algorithm: basic h-index to coreness}. As a bottom-up execution model, \emph{Peel} is simple and intuitive, but the inherently sequential processing requires the global graph information, leading to a fixed minimum number of iterations. As a top-down execution model, \emph{Index2core} can be formulated using the vertex-centric parallel model, but the total computational overhead is heavier than the \emph{Peel}-based algorithms in the shared-memory setting.

\begin{algorithm}[t]
    \caption{Peel paradigm} \label{Core-Number-peel} %
    \KwIn{$G_{res} \leftarrow G$; $k \leftarrow 0$; $core[v]\leftarrow 0$ for $ v \in V(G)$\;}
    \While{$G_{res} \neq \varnothing$}{
        $k$++\;   
        \While{$\exists v \in V(G_{res}): deg(v, G_{res}) <= k$}{
            $core[v] \leftarrow k$\;
            $v \leftarrow$ removed\; 
            for $u \in nbr(v,G_{res})$: $deg(u)$ - -\;
        }
    }
    \KwRet $core$\;
\end{algorithm}  

\SetKwProg{Fnn}{Function}{:}{}
\SetKwFunction{HINDEX}{HINDEX}
\begin{algorithm}[b]
    \caption{Index2core paradigm} 
    \label{algorithm: basic h-index to coreness}%
     \KwIn{$G$; $core[v] \leftarrow deg(v)$ $\forall v \in V(G)$\;}
     \Repeat{$core$ no longer changes}{
        \For{$v \in V$} {  
       $core[v] \leftarrow$ \HINDEX{$nbr(v)$, $core[]$}\;
       }
    }
 \Fnn{\HINDEX{$nbr(v)$, $core[]$}}{
     Return an integer $h$: \\
     $|u \in nbr(v): core[u] \geq h| \geq h $\\
     \& $|u \in nbr(v): core[u] \geq h+1| \leq h $
    
    }
\end{algorithm}

\subsection{Graph On GPU}

In 2001, NVIDIA designed the first graphics processing unit (GPU), deploying it to accelerate image and video processing applications. In 2006, NVIDIA released the first Tesla architecture-based GPU, extending support beyond visual processing to include scientific computing applications. Harish first introduced the Tesla architecture-based GPU to accelerate graph algorithms in 2007~\cite{graph-gpu-first}. Modern GPUs provide a massively parallel computing capability with thousands of threads. The impressive computing power requirements of widely used applications have driven GPUs to be the general-purpose computing co-processors in heterogeneous computers. Researchers from both industry and academia have leveraged GPUs to accelerate a variety of graph algorithms and graph processing frameworks~\cite{bfs, tc, sssp, gunrock, tigr1}. 

However, there are still great challenges to unleashing the massive parallelism capability of modern GPUs due to the irregular structure and random memory access patterns of real-world graphs.
The GPU-based k-core decomposition framework includes several hardware optimization strategies tailored specifically to leverage the architecture of GPUs.

\subsubsection{Graph Storage}
Dealing with large-scale graphs that encompass millions or even billions of vertices and edges is a significant challenge, particularly when considering the limited memory capacity of GPU accelerators. In order to solve this problem, most existing works have implemented the compressed sparse row (CSR) format to load the graph topology. A CSR compacts all the entities into two arrays. One is used to keep the sequential concatenation of the neighbors each vertex, while the second one is used to store the start location of the neighbor list of each vertex. While this format enables immediate neighbor identification for any specified vertex, it also leads to random memory access, thereby impeding GPU efficiency.

\subsubsection{Programming Models} 
To navigate the challenges of aligning graph vertices or edges with GPU threads, researchers have pioneered the vertex-centric~\cite{vertex-centric} and edge-centric~\cite{edge-centric} programming models to deploy the graph algorithms on GPUs. In the vertex-centric programming model, each vertex is assigned to a GPU thread/warp, simplifying algorithm implementation for developers. However, this model will lead to load imbalance issues due to the irregular graph structure of the real-world graphs. In the edge-centric programming model, each edge but not vertex is assigned to a GPU thread/warp, offering a natural solution to load imbalance. Nevertheless, this approach increases the computational workload because there are many more edges than vertices in sparse graphs. In pursuit of fully exploiting GPU capabilities, an array of innovative programming models has been proposed, including the path-centric~\cite{path-centric}, sub-graph-centric~\cite{graph-centric}, and data-centric~\cite{gunrock} programming models, etc.

\subsubsection{Thread Mapping} 
Thread mapping is a set of techniques that map the GPU thread/warp to a graph task. The basic method involves mapping a vertex or edge to a GPU thread/warp in conjunction with the vertex/edge-centric programming model. In order to improve GPU thread efficiency, the virtual warp and some other new techniques have been proposed in the state-of-the-art research~\cite{virtual-warp, tigr1}, which can allocate unfixed threads/warps to the vertex/edge according to the computing requirements of different tasks.


\subsection{Motivation}
\label{Motivation}

In this subsection, we first analyze the major challenges of the \emph{Peel}-based k-core decomposition paradigm, and then we conduct experiments on an RTX 3090 GPU to explore the performance bottlenecks of the \emph{Index2core} paradigm.

\begin{figure}[t]
    \centering
    \includegraphics[width=7.2cm]{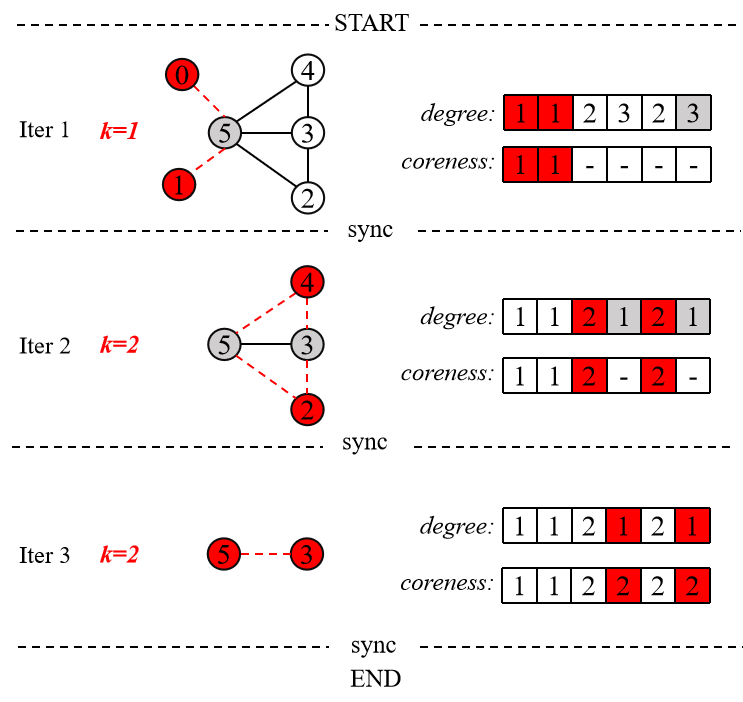}
    \caption{The commonly utilized parallel \emph{Peel} method in parallel.}
    \label{fig:peel-paradigm-existing}
\end{figure}

When locating the $k$-core within the \emph{Peel} paradigm, the removal of some vertices updates the degrees of residual vertices, which may fall below the specified coreness $k$. To facilitate subsequent iterations, these vertices must undergo additional processing to return to the correct coreness, which inevitably complicates the parallel logic of the entire algorithm and deteriorates performance. A commonly utilized method~\cite{GPU-core1, GPU-core3, gunrock} employs two distinct property arrays to independently track the residual degree and coreness, recognizing that these values may diverge. As illustrated in Fig.~\ref{fig:peel-paradigm-existing}, during the third iteration, the degree of $v_3$ and $v_5$ decrease to 1, which is below the coreness of 2. In Algorithm~\ref{Algo:naive peel}, the \emph{Peel} algorithm consists of two parallel graph operations with complex judgment conditions: a \emph{scan} operation to find frontiers, and a \emph{scatter} operation to update the degrees of the neighbors of these frontiers. In the \emph{scatter} kernel, the $atomicSub$ function may update some residual vertices once the coreness is below $k$. Since both removed vertices and those that remain have coreness values less than $k$, discerning whether a vertex has been removed based solely on the residual degree property is infeasible.
Thus, an additional flag, $rem$, is needed additional to ensure that only the degrees of the residual vertices are updated. 
In the \emph{scan} kernel, the criteria for identifying frontier nodes are multifaceted, requiring checks on both $rem$ and $deg$ variables. Another method~\cite{GPU-core4} appends an increased number of atomic add operations on these vertices, which causes serious atomic competition. 

\SetKwProg{Fn}{Kernel}{:}{}
\SetKwFunction{Fkernel}{scan}
\SetKwFunction{Scatterkernel}{scatter}
\begin{algorithm}[t]
    \caption{General Parallel Peel on GPU} \label{Algo:naive peel}
       $k \gets 0$; $rem[v] \leftarrow 0$ $\forall v \in V(G)$\;
    \While{$|V_{res}| > 0$}{
    
        \uIf{$\forall v \in V_{res}, !(deg(v) \leq k)$}
        {k++;}
        
        \Fn{\Fkernel{$G$, $rem[]$, $deg[]$, $core[]$, $k$}}
        {
        Frontiers:\\$V_f=\{v: !rem[v] \& deg[v] \leq k | v \in V(G)\}$;\\
        set $core[v] = k$, $v\in V_f$;\\
        set $rem[v] = true$, $v\in V_f$;\\
        }
        \Fn{\Scatterkernel{$G$, $rem[]$, $deg[]$}}
        {
            Graph Operator on the frontier:\\
            \For{$u \in nbr(v)$}
            {
                \uIf{{$!rem[u]$}}
                {
                $atomicSub(deg[u], 1)$;
                }
            }
        }

    }

\end{algorithm}

\begin{figure}[b]
    \centering
    \includegraphics[width=8cm]{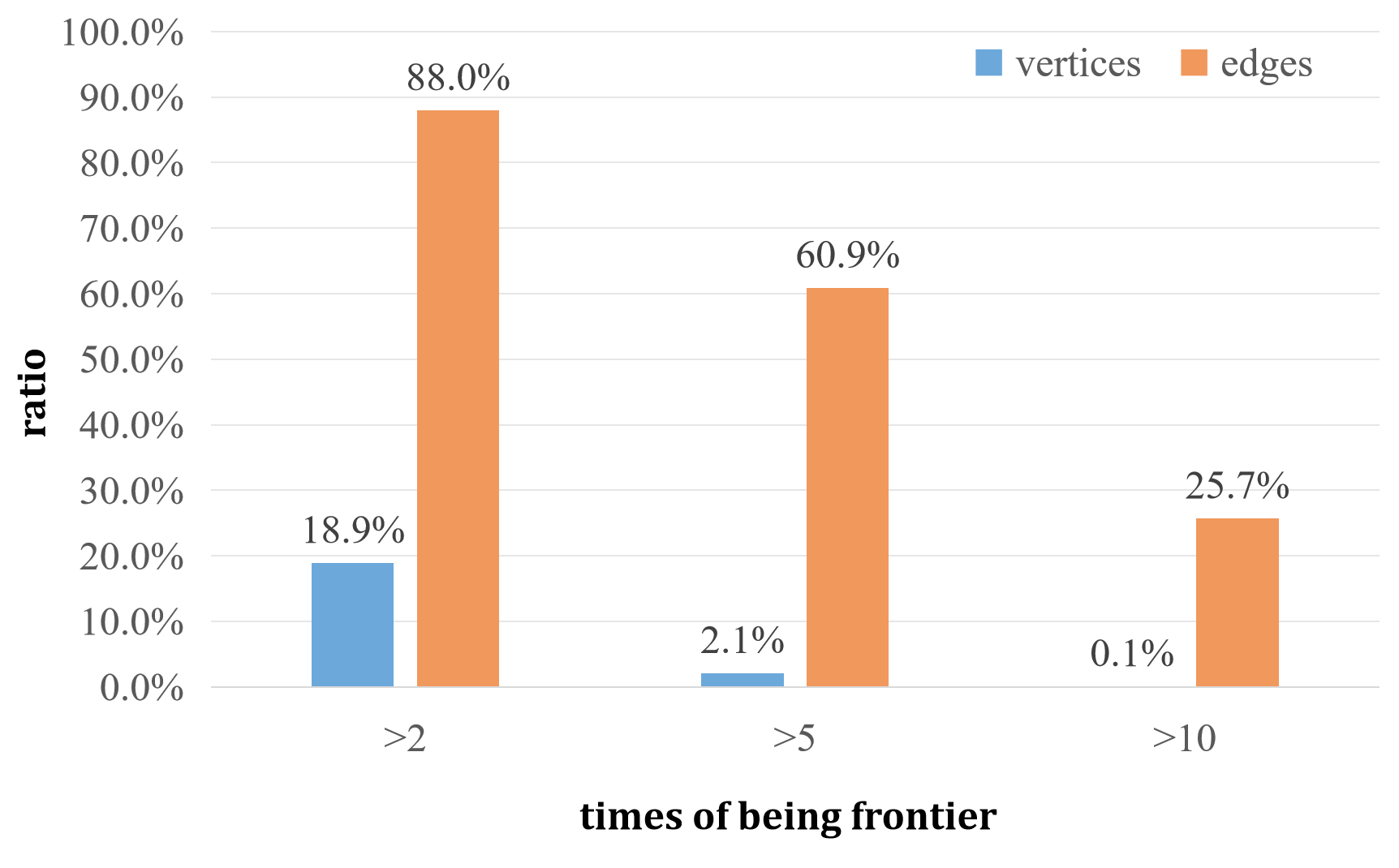}
    \caption{The proportion of vertices and edges that need multiple access in dataset \emph{soc-twitter-2010}.}
    \label{fig:index2core-motivation}
\end{figure}


To explore the bottlenecks of the \emph{Index2core} paradigm on GPU, we designed an experiment on RTX 3090 GPU to collect the frontiers of the power-law graph \emph{soc-twitter-2010} (the detailed experimental settings are the same as the experiments in Section \ref{exp}). The experimental result is shown in Fig.~\ref{fig:index2core-motivation}. From this experiment, we can first conclude that, for a fixed iteration, only a few neighbors of the frontier are required to participate in the next iteration. In the given example graph $G_1$, when the $h$-index value of $v_5$ changes from 5 to 2, it is not necessary to re-estimate the coreness of $v_0$, $v_1$, $v_2$ and $v_4$. Our experiment found that the $h$-index of an average of 94\% of the frontiers' neighbors stays unchanged. Another observation is that some high-degree vertices may become the frontiers more than once so that the estimation changes multiple times. Thus, another time-consuming operation is that the neighbors of those multi-changed frontiers are accessed multiple times across different iterations. Fig.~\ref{fig:index2core-motivation} shows that 18.9\% of vertices can be the frontier more than 2 times, while as many as 88\% of edges will be accessed more than 2 times and 60.9\% will be accessed more than 5 times. Based on the two observations, many redundant computations exist in the parallelizing \emph{Index2core} paradigm.

%% file: sec3_peelone.tex
\section{Design of the Proposed PeelOne Algorithm}

\label{peelone}

This section first introduces the definition of the under-core vertex and then proposes the \emph{PeelOne} algorithm with a low-cost method to eliminate the under-core vertex.

\subsection{The Under-Core Vertex}

In the \emph{Peel}-based $k$-core decomposition paradigm, neighbor vertices are removed iteratively, resulting in some active vertex with a degree lower than the coreness. 
We use $deg_r(u, k, G)$ to represent the residual degree of vertex $u$ when locating the $k$-core.
To effectively deal with these vertices, we introduce the concept of \emph{under the coreness} and define \emph{under-core}.

\begin{defi}[Under-Core Vertex]\label{def:under-core}
    For a given graph $G=(V,E)$, $u \in V$ is a under-core vertex if and only if $deg_r(u, k, G) < core(u, G)$.
\end{defi}

According to Definition \ref{def:under-core}, we further express the under-core vertex set as $V_{<core}(G)=\{u: deg_r(u, k, G) < core(u, G)| u \in V(G)\}$.

We illustrate the \emph{Peel} paradigm of $k$-core algorithm on the example graph $G_1$ in Fig.~\ref{fig:peel-paradigm-existing}. Three iterations are needed to locate the coreness of all the vertices in $G_1$. In the first two iterations, the degree of vertices $v_0$, $v_1$, $v_2$, and $v_4$ are equal to their coreness. However, the degree of the residual vertices $v_3$ and $v_5$ are less than the coreness in the third iteration. Thus, $v_3$ and $v_5$ are the under-core vertices and we can conclude that $V_{<core}(G1) = \{v_3, v_5\}$.

\subsection{The $assertion$ Method}

When accelerating the \emph{Peel} algorithm on GPU, the under-core vertices must undergo additional atomic processing to return the correct coreness, which inevitably complicates the parallel logic of the entire algorithm and reduces the performance. This subsection introduces a key theorem of under-core vertices that can avoid additional processing.

\begin{theo}\label{theo:under-core}
When locating the $k$-core in the $G$, the coreness of the under-core vertex is $k$.
\end{theo}

\begin{IEEEproof}
    When locating the $k$-core, the coreness of the removed vertices must be $\leq$ $k$. Thus, the coreness of the residual vertices must be $\geq$ $k$. If the degree of a vertex $u$ of the residual vertices is less than $k$, the vertex can not belong to the $k$+1-core of $G$. Thus, the coreness of $u$ is $k$ and $u$ is an under-core vertex.
\end{IEEEproof}

Theorem~\ref{theo:under-core} shows that the coreness of the under-core vertex equals $k$ while locating the $k$-core. Thus, when updating the degree of the frontiers' residual neighbors, we directly assign a coreness of $k$ to an under-core vertex rather than reducing their coreness below $k$. 
We design the method named $assertion$, including a novel atomic operation $atomicSub_{\geq k}(*address, 1, k)$. This operation performs the following steps: it reads the old value $old$ located at the address $address$ in global or shared memory, computes $(old > k) ? (old-1): k)$, and then stores the result back in memory at the same address. These three steps are executed in a single atomic transaction.

\begin{figure}[t]
    \centering
    \subcaptionbox{The workflow of the \emph{atomicAdd} method\label{fig:atomic-opt-general}}{
    \includegraphics[scale=0.45]{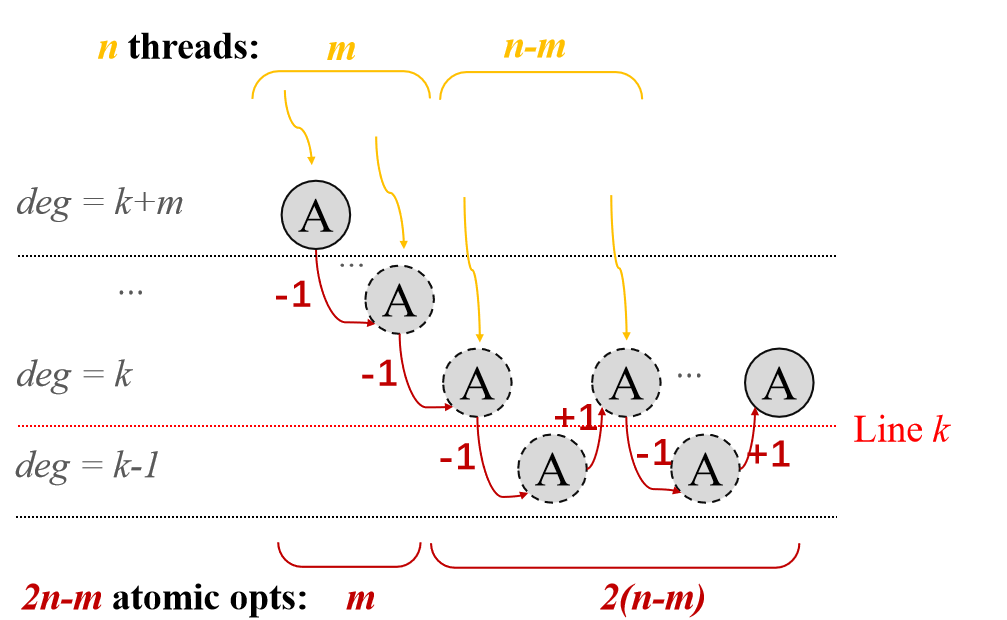}
    }
        \quad
    \subcaptionbox{The workflow of the proposed $assertion$ method
    \label{fig:atomic-opt-rollback}}{
    \includegraphics[scale=0.45]{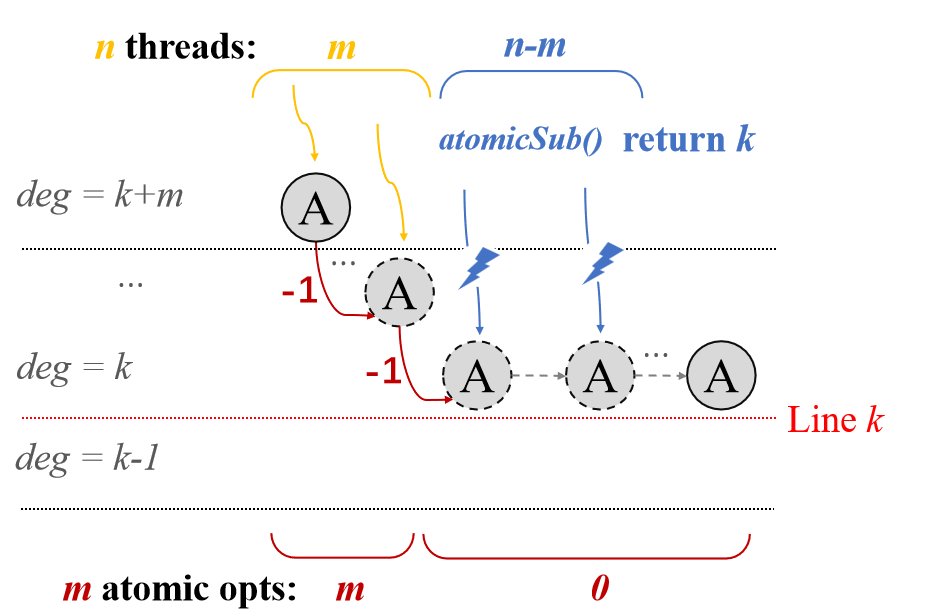}
    }
    \caption{The atomic operations involved in the reduction of the degree of under-core vertices.}
    \label{fig:atomic-opt}
\end{figure}

We show the general situation of processing an under-core vertex using an extra atomic add function in Fig.~\ref{fig:atomic-opt-general}. Supposing the $k$-core is located, the residual degree of vertex $A$ is $k+m$, and there are $n$ ($n > m$) threads (neighbors) performing an atomic minus one operation on vertex $A$. There are $n-m$ threads that will reduce the degree below k, and then the $n-m$ threads will atomically add one on $A$ to ensure that the residual degree is equal to the current $k$. Thus, a total of $2n-m$ atomic operations are performed to obtain the coreness of vertex $A$. Fig.~\ref{fig:atomic-opt-rollback} shows the workflow of our $assertion$ method. Ideally, the degree of $A$ is reduced to $k$, and then no more atomic operations are performed so that there are the additional $2(n-m)$ atomic operations avoided. In summary, our $assertion$ method reduces the redundant atomic operations and improves the performance by avoiding more atomic competition.

\subsection{The Proposed PeelOne Algorithm}

With the $assertion$ method, we can derive \emph{Corollary}~\ref{corolary: peel-filter}, which means that the degree of the residual vertices is not less than $k$ while locating the $k$-core. 

\begin{corl}
\label{corolary: peel-filter}
    When locating $k$-core in \emph{Peel} with the $assertion$ method, $\forall u \in V_{res}$, we have $deg(u, G_{res})\geq k$.
\end{corl}
\begin{IEEEproof}
     When locating $k$-core in \emph{Peel}, the coreness of the residual vertices is $\geq k$. Since we eliminate the under-core vertices, the degree of the residual vertices is $\geq k$.
\end{IEEEproof}

Therefore, the parallel logic of \emph{Peel} can be optimized in three steps as shown in \emph{PeelOne} Algorithm~\ref{k-core-Peel-one}.

\begin{algorithm}[b]
\caption{PeelOne} \label{k-core-Peel-one}
        $k \gets 0$; $core[v] \gets deg[v]$ $\forall v \in V(G)$\;
    \While{$|V_{res}| > 0$}{
             $k$++\;
        \Fn{\Fkernel{$G$, $core[]$, $k$}}
        {
        Frontiers:{$V_f= \{v: core[v] = k | v \in V(G)\}$};\\
        }
        \Fn{\Scatterkernel{$G$, $core[]$}}
        {
            For all frontiers, do Graph operator:\\ 
            \For{$u \in nbr(v)$}
            {
                \uIf{$core[u] > k $}
                {
                $atomicSub_{\geq k}(core[u], 1, k)$;\\
                
                 \tcp{Dynamic Frontier}
                 \uIf{$core[u]=k$}{
                 add $u$ to the frontier;\\
                 } 
                }
            }
        }
    }
\end{algorithm}

(1) Asserting the frontiers as $\{ u | degree[u] = k, u\in V_{res} \}$. Since the degree of the residual vertices is not less than $k$, the vertices with a degree value of $k$ in the residual vertices can be asserted as the frontiers.

(2) The condition for a neighbor $u$ to perform graph operator is indicated as $core[u] > k$ (Line 10). This condition ensures that atomic functions are not applied to vertices that have been removed. Given that the degrees of all residual vertices are not less than $k$, and those with a degree exactly equal to $k$ are considered frontiers, the updating of degrees is restricted only to neighbors with degrees greater than $k$. Thus, the additional $delete$ flag is no longer required. The condition indicator and the graph operator can access the residual degree ($core[u]$) in the same address, which has a better data locality.

(3) Asserting under-core vertices as the next frontiers in advance. While doing the graph operator in the \emph{scatter}, the under-core vertices from the frontier's neighbors can be asserted the frontiers in the next iteration in advance. With the $assertion$ method, the neighbor $u$ with the $core[u]$ value of $k$ is, in fact, an ensuing frontier. The vertex $u$ with $core[u] = k$ can be accumulated in a dynamic frontier queue and processed within the current iteration. This approach eliminates the need for the synchronization overhead associated with the \emph{scan} and \emph{scatter} kernels in an additional iteration.

\begin{figure}[t]
    \centering
    \includegraphics[width=8.9cm]{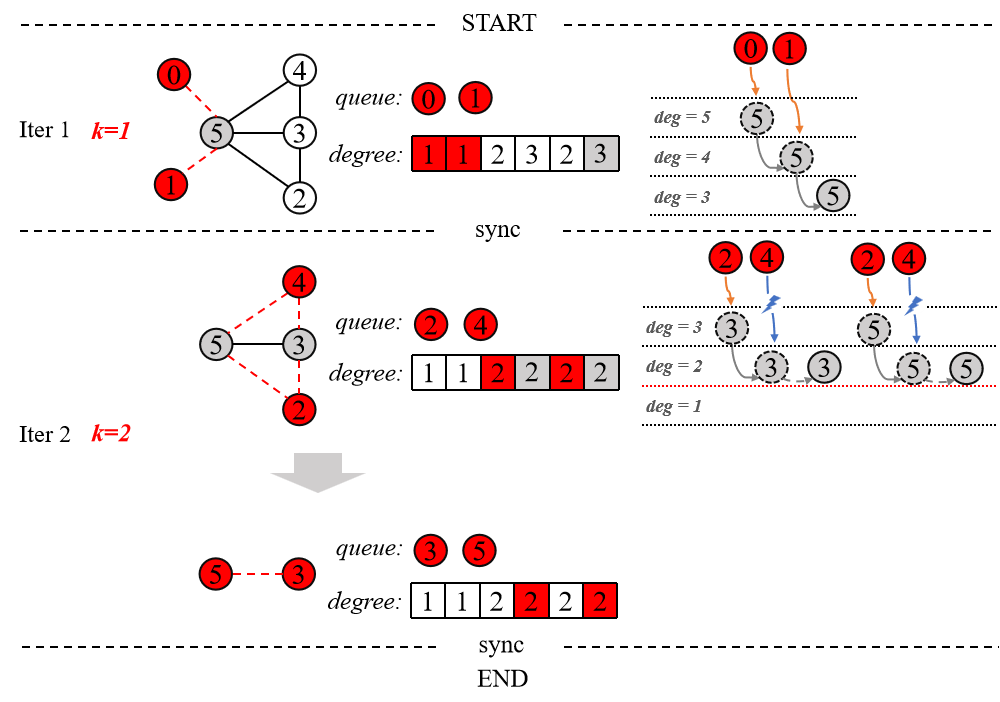}
    \caption{The procedure of \emph{PeelOne} method in parallel.}
    \label{fig:peel-paradigm-PeelOne}
\end{figure}

Fig.~\ref{fig:peel-paradigm-PeelOne} shows the details of \emph{PeelOne} on the example graph $G_1$. $v_2, v_4$ are the frontiers in the second iteration. When performing the graph operator on $v_2$ and $v_4$, the degree of $v_3$ and $v_5$ are large than $k$ ($k=2$) and then are updated to 2 by the function $atomicSub_{\geq k}$. Vertices $v_3$ and $v_5$ with a degree of $2$ are added to the frontier queue to perform the graph operator in this iteration.

%% file: sec4_hist.tex
\section{Design of the HistoCore Algorithm}
\label{hist}

This section describes our two proposed \emph{Index2core}-based algorithms. To reduce the redundant computation on vertices, we propose the \emph{CntCore} by precisely the frontiers in every iteration. Furthermore, to reduce the redundant computation on edges, we propose the \emph{HistoCore} by maintaining the up-to-date histogram information for the multi-changed vertices that become the frontiers many times.

\subsection{The Proposed CntCore}


Considering the \emph{Index2core} paradigm, since the coreness of some vertices can converge within a few iterations, it is important to determine the vertex set that participates in the next iteration of the computation. The existing method, in which the neighbors of the vertex that the $h$-index (coreness estimation value) is changed in the current iteration participate in the next iteration, has a large amount of redundant computation. Thus, we propose the \emph{CntCore} only to estimate the frontiers' coreness.

We denote the number of neighbors with an $h$-index value no smaller than vertex $u$ in the $t-1$-th iteration as $cnt(u, t) = |\{v: h^{t-1}_v \geq h^{t-1}_u, v \in nbr(u) \}|$. Thus, $cnt(u, t)$ represents the number of neighbors that could potentially influence the $h$-index of $u$ in the $t$-th iteration. The following theorem holds.

\begin{theo}
 \label{lemma: cnt}
   For a vertex $u \in G$, $h^{t}_u < h^{t-1}_u$ if and only if $ cnt(u, t) < h^{t-1}_u$.
\end{theo}

Theorem~\ref{lemma: cnt} demonstrates the necessary condition for determining whether the $h$-index of $u$ has changed ($h^{t}_u < h^{t-1}_u$) in the $t$-th iteration. Thus, in every iteration, only the $h$-index of the vertex set $V^{t}_{cnt}=\{cnt(u, t) < h^{t-1}_u, u \in V(G)\}$ are required to be estimated. In other words, the frontiers of every iteration in the \emph{Index2core} procedure are the set $V^{t}_{cnt}$.

\begin{algorithm}[t]
    \caption{CntCore}\label{Algo:cntCore}
    $V_{active} \gets V$;\\
     \While{$V_{active} \neq \varnothing$}{
      Compute the $cnt[v]$ in $V_{active}$;\\
      Frontiers:{$V_f= \{v: cnt[v] < core[v], v \in V_{active}\}$};\\
      $V_{active} \gets \varnothing$;\\
      Graph Operator:\\
      estimate the coreness by the call \HINDEX{$nbr(v), core[v]$} ;\\
      add the $nbr(v)$ set to the $V_{active}$;\\
    }
\end{algorithm}

Based on the above observation, we design the \textbf{\emph{CntCore}} as shown in the Algorithm~\ref{Algo:cntCore}. In the first iteration, compute the $cnt$ value of all the vertices, locate the frontiers with the $cnt$ value less than their degree (the initial coreness estimation value), and estimate the new value of the frontiers. To avoid computing the $cnt$ of all vertices in the next iteration, a set $V_{active}$ is used to store the union of all frontiers' neighbors. Then, in the next iteration, only the \emph{cnt} value in the $V_{active}$ are recomputed, and the frontiers are found from within $V_{active}$. In summary, the \emph{CntCore} estimates only the frontiers' coreness and avoids the redundant computation of vertices whose estimation values remain unchanged.

\subsection{The Proposed HistoCore}
\begin{figure}[b]
    \centering
    \includegraphics[width=8.5cm]{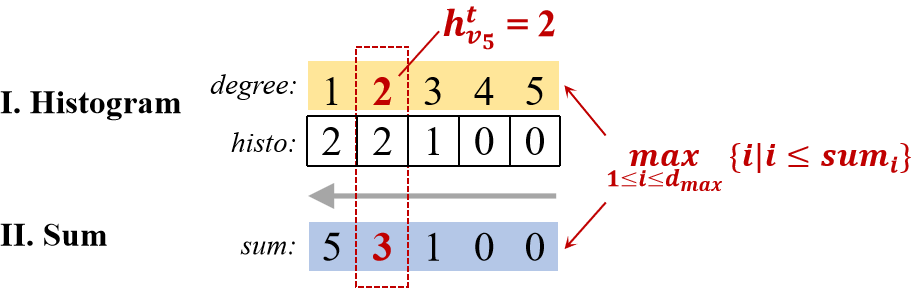}
    \caption{An example of calculating $h^t_{v_5}$ of vertex $v_5$.}
    \label{fig:computehindex}
\end{figure}

High-degree vertices might often become frontiers, meaning these vertices require calling the HINDEX function to estimate their coreness. To explore the redundant computations involved in the HINDEX function, we decouple the function into two steps: \emph{Step I: Histogram} and \emph{Step II: Sum}. As shown by the $h$-index example of vertex $v_5$ in Fig.~\ref{fig:computehindex}, \emph{Step I: Histogram} involves counting and storing the occurrences of each value among neighbors in the array $histo$ i.e., \{ $1:\{v_0, v_1\}$ $2:\{v_2, v4\}$ $3:\{v_3\}$\}. \emph{Step II Sum} involves computing the $h$-index value based on the cumulative sum in $histo$ by performing a reverse summation. With these two steps, the $h$-index of $v_5$ is determined to be 2. It is obvious that \emph{Step I: Histogram} leads to massive random memory access since it requires reading all neighbors' value and writing back to the $histo$ array. The \emph{Step II: Sum} operation only accesses a portion of the $histo$ array sequentially.

\begin{figure}[t]
    \centering
    \subcaptionbox{CntCore\label{fig:histo-array-cntCore}}{
    \includegraphics[scale=0.47]{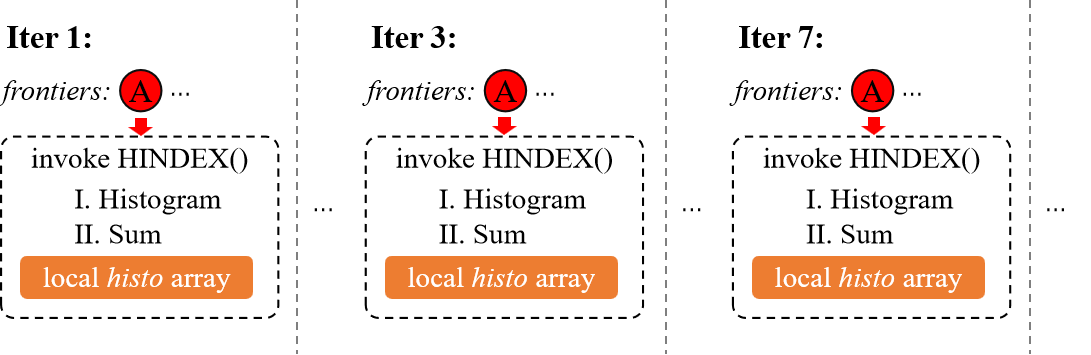}
    }
    \quad
    \subcaptionbox{HistoCore
    \label{fig:histo-array-histoCore}}{
    \includegraphics[scale=0.47]{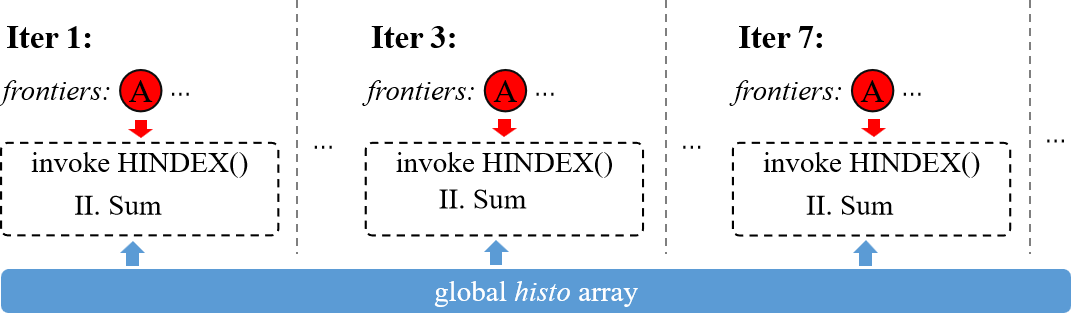}
    }
    \caption{The $histo$ array construction and maintenance of multi-changed vertices.}
    \label{fig:histo-array}
\end{figure}

Considering a high-degree vertex $A$ in the procedure of \emph{CntCore} as shown in Fig. \ref{fig:histo-array-cntCore}. In the 1st, 3rd, and 7th iterations, $A$ becomes the frontier for estimating its coreness using the HINDEX function. Therefore, the neighbors of $A$ are accessed a total of three times, and its $histo$ array is also constructed three times correspondingly. To reduce the time spent performing \emph{Step I: Histogram}, our approach is to maintain a global and up-to-date $histo$ array for $A$ so that $A$ only needs to perform \emph{Step II: Sum} when it becomes the frontier. Based on this idea, we propose the \emph{HistoCore} algorithm. Fig.~\ref{fig:histo-array-histoCore} shows the procedure of processing vertex $A$ with \emph{HistoCore}. $A$ can estimate its coreness by only performing the \emph{Step II: Sum} operation. Thus, the key issue of \emph{HistoCore} is to maintain the up-to-date $histo$ array.

\subsubsection{Maintain the Up-to-date $histo$ Array}

We analyze the effects of the different types of changes to the neighbor's $h$-index on a vertex's $h$-index. Supposing that the current $h$-index value of vertex $A$ is $\textbf{k}$, and the value of neighbor $N$ drops from $x$ to $y$ ($x > y$), we classify the neighbors with changed values into three types:

\begin{itemize}
\item Type N1: $\textbf{k} > x > y $
\item Type N2: $x > y \geq \textbf{k}$
\item Type N3: $x \geq \textbf{k} > y$
\end{itemize}

\begin{figure}[h]
    \centering
    \subcaptionbox{Original $histo$ array of vertex $A$\label{fig:neighbor-value-change-histo}}{
    \includegraphics[scale=0.7]{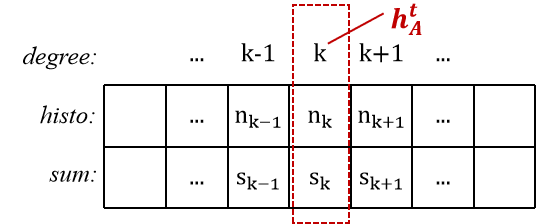}
    }
    \quad
    \subcaptionbox{Type N1: $k > x > y$
    \label{fig:neighbor-value-change-N1}}{
    \includegraphics[scale=0.7]{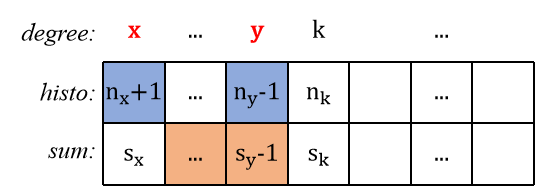}
    }
    \quad
    \subcaptionbox{Type N2: $x > y \geq k$
    \label{fig:neighbor-value-change-N2}}{
    \includegraphics[scale=0.7]{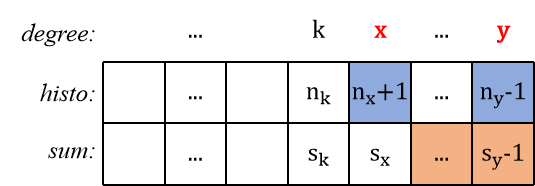}
    }
    \quad
    \subcaptionbox{Type N3: $x \geq k > y$
    \label{fig:neighbor-value-change-N3}}{
    \includegraphics[scale=0.7]{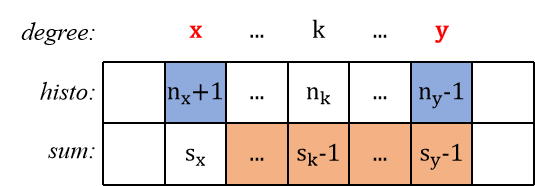}
    }
    \caption{Three situations of neighbor degree changes.}
    \label{fig:neighbor-value-change}
\end{figure}

We show an example that demonstrates the effect of these three types of neighbors on vertex $A$. 
We utilize $n_i$ to denote the number of neighbors with the $h$-index value of $i$, and $s_i$ to represent the sum of neighbors with a $h$-index no less than $i$, where $s_i=\sum^{d_{max}}_{i} n_i$. Given that the $h$-index of vertex $A$ is denoted as $\textbf{k}$, we can infer that $k \leq s_k$ and $k+1 > s_{k+1}$. As shown in Fig.~\ref{fig:neighbor-value-change}, N1 and N2 change certain values in the $histo$ and $sum$ arrays, which are highlighted in blue and orange, respectively. But the value of $s_k$ and $s_{k+1}$ remains unaffected, resulting in the unchanged $h$-index of $A$. In the N3 situation in Fig.~\ref{fig:neighbor-value-change-N3}, the change of neighbor degree from $y$ to $x$ causes a modification in $s_k$ and $s_{k+1}$, potentially resulting in a change of the $h$-index. Based on these observations, we draw the following three conclusions: 

\begin{description}
      \item[I.] Type N2 does not affect the $h$-index of vertex $A$;
      
      \item[II.] If Type N3 does not exist, Type N1 does not affect the $h$-index of vertex $A$;
      
      \item[III.] If Type N3 exists, Type N3 and Type N1 both affect the $h$-index of vertex $A$;
      
\end{description}

Thus, once the Type N1 or Type N3 exists, we update the $histo$ array of $A$. In conclusion, combining N1 and N3, \textbf{when the latest $h$-index value of a neighbor is less than that of $A$, we update the $histo$ array of $A$}. 


\subsubsection{Detailed Design of HistoCore }

We show the detailed pseudo-code in Algorithm~\ref{algorithm: HistoCore}. In the init kernel, we initialize a global $histo$ array for each vertex according to the degree of its neighbors. Then start the iteration with the $SumHisto$ kernel and the $UpdateHisto$ kernel.

In the $SumHisto$ kernel, we sum these up-to-date $histo$ arrays to get the latest $h$-index values of the frontiers. 
In the $UpdateHisto$ kernel, we assign threads to these vertices (the frontiers), whose values have changed in the last $SumHisto$ kernel, to update the $histo$ arrays of their neighbors.

Notably, the \emph{HistoCore} also finds the frontiers by the $cnt$ value but in a different way. We found that the $cnt$ value of a vertex can be included by the $histo$ array. In the $SumHisto$ kernel (Lines 10-15), there is a byproduct value, $sum$, in the procedure of calculating the $h$-index by summing the $histo$. This $sum$ essentially represents the $cnt$ value, so we store it in the address of the current value of the vertex in the $histo$. And in the $UpdateHisto$ kernel (Lines 20, 22-23), the $cnt$ value in the $histo$ array may be updated and can naturally be used to obtain the frontiers for the next iteration.

In summary, \emph{HistoCore} locates the frontiers to reduce invalid computations of the most active vertices and redundant access to their neighbors, thus \emph{HistoCore} can show the optimal parallelism of \emph{Index2core} paradigm.

\SetKwFunction{Fhistocore}{SumHisto}
\SetKwFunction{Fhistocnt}{UpdateHisto}
\SetKwFunction{FhistoInit}{InitHisto}
    \begin{algorithm}[hb]
    \caption{HistoCore}
    \label{algorithm: HistoCore} 
    $core[v] \gets deg[v]$ $\forall v \in V(G)$\;
    \Fn{\FhistoInit{$deg[]$}}{
     \For {$u \in nbr(v)$, $v \in V$} {
                        $histo[v][\min (core[u], core[v])]$++;
                     }
    }
    \While{$V_{cnt} \neq \varnothing$}{
    Frontiers: {v: same as CntCore}\\
    Graph operator on Frontiers :\\ sum and update the histo;\\
    \Fn{\Fhistocore{$histo[][]$, $core[]$, $oldcore[]$}}{
                        $core_{old} \leftarrow core[v]$; $sum \gets 0$\;
                     \For {$k \leftarrow$ $core_{old}$ to $ 1 $ s.t. $sum < k$} {
                        $sum \leftarrow sum + histo[v][k]$\;
                       
                     }
                      \uIf{$core_{old} \ne k$}{
                            $core[v] \gets k$; $oldcore[v] \leftarrow core_{old}$\;
                            \tcp{cnt[v] == sum;}
                            $histo[v][k] \gets sum$\;       
                     }
\KwRet \;
                    }
            \Fn{\Fhistocnt{$histo[][]$, $core[]$, $oldcore[]$}}{

                \For{$u \in nbr[v]$}{ 

                            \uIf{$core[u] > core[v]$}{

                                $ cnt\_value \gets$\\
                                $atomicSub(histo[u][min(oldcore[v], core[u])], 1)$;\\
                                $atomicAdd(histo[u][core[v]], 1)$\;
                                
                                \uIf{$oldcore[v] >= core[u]$ and $cnt\_value = core[u]$}
                                {
                                    add $u$ to the $V_{cnt}$;
                                }

           } 

           }

           }
    }
\end{algorithm}

%% file: sec5_exp.tex
\begin{table*}[ht]
\caption{Statistical properties of 24 datasets.}
\label{tab:datasets}
\setlength{\tabcolsep}{9pt}
\centering
\small
\begin{tabular}{ccccccccc}
\toprule
abridge & dataset          & $\left | V \right |$    &  $\left | E \right |$      & $d_{avg}$     & std     & $d_{max}$     & $k_{max}$ & Category          \\
\midrule
gow     & loc-Gowalla      & 197K     & 1,901K     & 9.6681   & 53.5776 & 14730    & 51   & Social Network    \\
ama     & amazon0601       & 403K     & 4,887K     & 12.1143  & 14.9456 & 2752     & 10   & Co-purchasing     \\
talk    & wiki-Talk        & 2,394K   & 9,319K     & 3.8921   & 102.508 & 100029   & 131  & Communication     \\
goo     & web-Google       & 916K     & 8,644K     & 9.4324   & 38.8326 & 6332     & 44   & Web Graph         \\
ber     & web-BerkStan     & 685K     & 13,299K    & 19.4080  & 285.162 & 84230    & 201  & Web Graph         \\
ski     & as-Skitter       & 1,696K   & 22,191K    & 13.0809  & 136.861 & 35455    & 111  & Internet Topology \\
pat     & cit-Patents      & 3,775K   & 33,038K    & 8.7523   & 10.4908 & 793      & 64   & Citation Network  \\
in      & in-2004          & 1,383K   & 27,183K    & 19.6564  & 146.566 & 21869    & 488  & Web Graph         \\
dbl     & dblp-author      & 5,624K   & 24,564K    & 4.3676   & 10.527  & 1389     & 14   & Collaboration     \\
woc     & wikipedialink-oc & 96K      & 29,273K    & 304.2076 & 858.598 & 40619    & 1252 & Web Graph         \\
lj      & LiveJournal1     & 4,848K   & 85,702K    & 17.6795  & 52.0034 & 20333    & 372  & Social Network    \\
wde     & wikipedialink-de & 3,604K   & 155,094K   & 43.0371  & 497.892 & 434234   & 837  & Web Graph         \\
hol     & hollywood-2009   & 1,140K   & 112,751K   & 98.9145  & 271.867 & 11467    & 2208 & Collaboration     \\
ork     & com-Orkut        & 3,072K   & 234,370K   & 76.2814  & 154.781 & 33313    & 253  & Social Network    \\
tra     & trackers         & 27,666K  & 281,227K   & 10.1652  & 2773.95 & 11571953 & 438  & Web Graph         \\
ind     & indochina-2004   & 7,415K   & 301,970K   & 40.7249  & 390.704 & 256425   & 6869 & Web Graph         \\
uk      & uk-2002          & 18,520K  & 523,575K   & 28.2702  & 144.861 & 194955   & 943  & Web Graph         \\
sina    & soc-sinaweibo    & 58,656K  & 522,642K   & 8.9103   & 165.497 & 278489   & 193  & Social Network    \\
twi     & soc-twitter-2010 & 21,298K  & 530,051K   & 24.8876  & 414.072 & 698112   & 1695 & Social Network    \\
wien    & wikipedialink-en & 13,593K  & 669,183K   & 49.2299  & 644.587 & 1052326  & 1114 & Web Graph         \\
ara     & arabic-2005      & 22,744K  & 1,107,806K & 48.7075  & 555.208 & 575628   & 3247 & Web Graph         \\
uk      & uk-2005          & 39,460K  & 1,566,054K & 39.6872  & 1655.12 & 1776858  & 588  & Web Graph         \\
wb      & webbase-2001     & 118,142K & 1,709,620K & 14.4709  & 143.961 & 816127   & 1506 & Web Graph         \\
it      & it-2004          & 41,291K  & 2,054,950K & 49.7671  & 883.439 & 1326744  & 3224 & Web Graph   \\
\bottomrule
\end{tabular}
\end{table*}

\section{EXPERIMENTAL EVALUATIONS}
\label{exp}
In this section, we evaluate the performance of all the proposed algorithms. To avoid the overhead from high-level programming, the proposed algorithms and baselines are all implemented by the low-level programming language CUDA C++. Furthermore, to reduce the load imbalance issues, we adopt a generally accepted technique~\cite{tigr1} for our proposed algorithms.

\subsection{Setup}

\subsubsection{Environments}
All algorithms are implemented in CUDA C++ and evaluated on an NVIDIA GeForce RTX 3090 GPU with 24GB of global memory and  10496 CUDA Cores. The CUDA Driver Version used is 12.2. The detailed parameters of the RTX 3090 GPU are shown in Table~\ref{tab:gpu}. For fairness, we run all the algorithms for 20 times and report the average time(milliseconds).

\begin{table}[b]
\caption{The main performance parameters of GPU.}
\label{tab:gpu}
\setlength{\tabcolsep}{20pt}
\centering
\small
\begin{tabular}{cc}
 \toprule
Device Name         & GeForce RTX 3090 \\ 
\midrule
CUDA Driver Version & 12.2 \\
CUDA Capability     & 8.6   \\ 
Global Memory:      & 24G   \\ 
Multiprocessors:    & 82    \\
CUDA Cores/SM:      & 128  \\
Maximum threads per SM: &1536 \\
GPU Max Clock rate: & 1725 MHz \\   
Memory Clock rate:  & 9751 Mhz  \\ 
Memory Bus Width:   & 384-bit   \\  
\bottomrule
\end{tabular}
\label{table:GPU}
\end{table}

\subsubsection{Datasets}

We conduct experiments using our GPU algorithms and other baselines on 24 publicly available datasets that vary in category, size, average vertex degree, maximum vertex degree and the max coreness, as shown in Table~\ref{tab:datasets}. These datasets are sourced from different categories:
\begin{itemize}
    \item Web graphs like \emph{web-Google}, \emph{web-BerkStan}, \emph{trackers}, \emph{webbase-2001}, \emph{uk-2002}, \emph{it-2004}, \emph{in-2004}, \emph{indochina-2004}, \emph{arabic-2005}, \emph{uk-2005}, \emph{wikipedia-link-oc}, \emph{wikipedia-link-de}, \emph{wikipedia-link-en}.
    \item Interaction networks such as communication network \emph{wiki-Talk}, citation network cit-Patents, collaboration networks \emph{dblp-author} and \emph{hollywood-2009}, as well as social networks \emph{com-Orkut}, \emph{loc-Gowalla}, \emph{soc-LiveJournal1}, \emph{soc-sinaweibo} and \emph{soc-twitter-2010}.
    \item Internet topology network \emph{as-Skitter}.
    \item Co-purchasing network \emph{amazon0601}.
\end{itemize}

These datasets have been widely used in previous related studies. As observed, the graphs demonstrate diverse characteristics. The number of vertices (respectively edges) reaches up to 118.14 million on \emph{webbase-2001)} (respectively 2.05 billion on \emph{it-2004}). The degree distribution exhibits a significant bias on \emph{trackers}, with an average degree of only 10.17, yet boasting a maximum vertex degree as high as 1.16 million. The max coreness reaches up to 6,869 on \emph{Indochina-2004}.

\subsubsection{Baselines}
The following shows the baselines and the proposed algorithms.
\begin{itemize}
    \item \emph{General Parallel Peel (GPP)}: This algorithm is implemented based on the work~\cite{GPU-core1, GPU-core3} and the $k$-core implementation in GPU graph system Gunrock~\cite{gunrock}.
    \item \emph{PeelOne}: The \emph{PeelOne} is proposed in this paper.
    \item \emph{Parallel Peel with the dynamic frontier (PP-dyn)}: This is the latest and SOTA work of \emph{Peel} on GPU~\cite{GPU-core4}, which is implemented by low-level programming with a finely optimized block-level dynamic frontiers queue. They perform extra atomic operations to process the under-core vertices. 
    \item \emph{PeelOne with dynamic frontiers (PO-dyn)}: The proposed \emph{PeelOne} algorithm combined with the dynamic frontier method and the $assertion$ method.
    \item \emph{NbrCore}: This \emph{Index2core}-based algorithm on GPU is implemented based on the work~\cite{GPU-core1}. It recomputes all neighbors' values of a vertex if the value of the vertex changes.
    \item \emph{CntCore}: This \emph{Index2core}-based algorithm is proposed in this paper. We locate the frontiers by the \emph{cnt} value to reduce the redundant computation on vertices.
    \item \emph{HistoCore}: This \emph{Index2core}-based algorithm proposed in this paper. We maintain the up-to-date \emph{histo} array for the vertex to reduce the redundant accessing on edges.
\end{itemize}

\begin{table}[hb]
\caption{The execution time of \emph{GPP} and \emph{PeelOne}.}
\setlength{\tabcolsep}{8pt}
\centering
\small
\label{tab:time-peel}
\begin{tabular}{cccccc}
\toprule
Dataset & GPP     & PeelOne     & Speed Up  & Gunrock & $l_1$ \\
\midrule
gow     & 25.2    & 21.0    & 1.2$\times$      & 58.3772                     & 647   \\
ama     & 10.5    & 8.3     & 1.3$\times$      & 21.4289                     & 258   \\
talk    & 67.8    & 40.8    & 1.7$\times$      & 476.841                     & 812   \\
goo     & 27.4    & 18.7    & 1.5$\times$      & 54.8884                     & 428   \\
ber     & 112.5   & 89.1    & 1.3$\times$      & 439.928                     & 2519  \\
ski     & 97.2    & 63.3    & 1.5$\times$      & 548.144                     & 1306  \\
pat     & 119.9   & 60.7    & 2.0$\times$      & 421.405                     & 1017  \\
in      & 193.9   & 134.0   & 1.4$\times$      & 2116.87                     & 3351  \\
dbl     & 27.2    & 12.7    & 2.1$\times$      & 78.2909                     & 183   \\
woc     & 119.6   & 114.7   & 1.0$\times$      & 5292.7                      & 3084  \\
lj      & 464.1   & 244.4   & 1.9$\times$      & 3107.48                     & 3851  \\
wde     & 532.9   & 328.4   & 1.6$\times$      & 11415.7                     & 4386  \\
hol     & 562.4   & 414.5   & 1.4$\times$      & 25285.2                     & 7462  \\
ork     & 772.5   & 541.4   & 1.4$\times$      & 5728.11                     & 5919  \\
tra     & 1581.2  & 417.6   & 3.8$\times$      & 120380                      & 3032  \\
ind     & 3585.6  & 1825.5  & 2.0$\times$      & 234218                      & 20180 \\
uk      & 3571.8  & 1782.1  & 2.0$\times$      & 35607.9                     & 9461  \\
sina    & 3238.7  & 783.4   & 4.1$\times$      & 9211.47                     & 3103  \\
twi     & 4965.7  & 1958.8  & 2.5$\times$      & 106421                      & 11436 \\
wien    & 2985.7  & 1413.1  & 2.1$\times$      & 52981                       & 8514  \\
ara     & 12773.6 & 6756.1  & 1.9$\times$      & 312103                      & 24951 \\
uk      & 8355.0  & 4223.6  & 2.0$\times$      & \multicolumn{1}{r}{OOM}     & 10143 \\
wb      & 47269.5 & 20279.5 & 2.3$\times$      & \multicolumn{1}{r}{OOM}     & 22814 \\
it      & 36176.7 & 20330.9 & 1.8$\times$      & \multicolumn{1}{r}{OOM}     & 38813 \\
\bottomrule
\end{tabular}
\end{table}

\subsection{Results and Discussion}
Our experiment focuses on four main objectives, as follows:
\begin{itemize}
    \item We aim to examine the performance of the \emph{PeelOne} algorithm in comparison with the \emph{General Parallel Peel} algorithm.
    \item We investigate whether the latest \emph{Peel} on GPU can be further improved by following the guidelines of the \emph{PeelOne} algorithm.
    \item We analyze the performance of \emph{HistoCore} in the context of the \emph{Index2core} paradigm and compare it with the performance of \emph{NbrCore} and \emph{CntCore}.
    \item We compare the performance of the optimal \emph{Peel} paradigm and the optimal \emph{Index2core} paradigm.
\end{itemize}

\subsubsection{Parallel Peel-based Algorithm Time}
\textbf{\emph{PeelOne} achieves an average speedup of 1.9$\times$ compared to \emph{GPP}}. As previously mentioned, \emph{PeelOne} simplifies the logic of the frontiers and the graph operators with the $assertion$ method. For the frontiers, \emph{PeelOne} reduces the usage of the property array and merges the residual degree and the coreness of the vertex in one array. For the graph operators, the residual degree can be used as the condition of the residual vertex so that the memory accesses are localized. Table~\ref{tab:time-peel-without} illustrates the advantages of the proposed \emph{PeelOne}. The result shows that \emph{PeelOne} achieves up to 4.1$\times$ on the \emph{soc-sinaweibo} and the average speedup is 1.9$\times$ on the 24 datasets. In addition, we also show the execution time of the implementation in the graph-parallel GPU system \emph{gunrock}. Due to system-level overheads, this implementation is clearly slower than \emph{GPP} and \emph{PeelOne}.

\textbf{The \emph{PeelOne} method can dramatically improve its performance by supporting dynamic frontiers and the $assertion$ method}. As shown in the Table~\ref{tab:time-peel-without}, $l_1$ is the number of iterations of the algorithms. By supporting dynamic frontiers, the $l_1$ equals the max coreness of the dataset, and the $l_1$ is significantly reduced (2$\times$$\sim$25.8$\times$, average: 11$\times$). As a result, the overhead of synchronization and locating the frontiers can be drastically reduced. Thus, \emph{PP-dyn} achieves an average speedup of 5.2$\times$. Furthermore, with the proposed $assertion$ method, 
\emph{PeelOne-dyn} achieves optimal performance by avoiding more atomic competition.
In summary, \emph{PeelOne} with the dynamic frontier can achieve the best performance in the \emph{Peel}-based algorithms. 


\begin{table}[h]
\caption{The performance of \emph{PeelOne} with dynamic frontiers and assertion method.}
\setlength{\tabcolsep}{4pt}
\centering
\small
\label{tab:time-peel-without}
\begin{tabular}{ccccc}
\toprule
Dataset & PeelOne ($l_1$) & PP-dyn ($l_1$) & Speed Up & \textbf{PO-dyn ($l_1$)} \\
\midrule
gow              & 21(647)        & 3(51)            & 7$\times$       & \textbf{3}(51)                \\
ama              & 8.3(258)       & 1(10)            & 8.3$\times$     & \textbf{1}(10)                \\
talk             & 40.8(812)      & 25(131)          & 1.6$\times$     & \textbf{24}(131)              \\
goo              & 18.7(428)      & 3(44)            & 6.2$\times$     & \textbf{3}(44)                \\
ber              & 89.1(2519)     & 15.3(201)        & 5.8$\times$     & \textbf{14.8}(201)            \\
ski              & 63.3(1306)     & 23.4(111)        & 2.7$\times$     & \textbf{22.9}(111)            \\
pat              & 60.7(1017)     & 10(64)           & 6.1$\times$     & \textbf{10}(64)               \\
in               & 134(3351)      & 25(488)          & 5.4$\times$    & \textbf{22}(488)              \\
dbl              & 12.7(183)      & 7(14)            & 1.8$\times$    & \textbf{7}(14)                \\
woc              & 114.7(3084)    & 54(1252)         & 2.1$\times$    & \textbf{59.8}(1252)           \\
lj               & 244.4(3851)    & 58.9(372)        & 4.2$\times$    & \textbf{56.7}(372)            \\
wde              & 328.4(4386)    & 216.1(837)       & 1.5$\times$    & \textbf{211}(837)             \\
hol              & 414.5(7462)    & 150.9(2208)      & 2.7$\times$    & \textbf{136.7}(2208)          \\
ork              & 541.4(5919)    & 107.9(253)       & 5$\times$      & \textbf{104}(253)             \\
tra              & 417.6(3032)    & 1032.6(438)      & 0.4$\times$    & \textbf{1030.8}(438)          \\
ind              & 1825.5(20180)  & 565.9(6869)      & 3.2$\times$    & \textbf{514.7}(6869)          \\
uk               & 1782.1(9461)   & 213.1(943)       & 8.4$\times$    & \textbf{207.3}(943)           \\
sina             & 783.4(3103)    & 471.7(193)       & 1.7$\times$    & \textbf{467.6}(193)           \\
twi              & 1958.8(11436)  & 918.9(1695)      & 2.1$\times$    & \textbf{914.2}(1695)          \\
wien             & 1413.1(8514)   & 693.3(1114)      & 2$\times$      & \textbf{690.1}(1114)          \\
ara              & 6756.1(24951)  & 889.6(3247)      & 7.6$\times$    & \textbf{869.2}(3247)          \\
uk               & 4223.6(10143)  & 449.7(588)       & 9.4$\times$    & \textbf{437.7}(588)           \\
wb               & 20279.5(22814) & 1396.7(1506)     & 14.5$\times$   & \textbf{1387.2}(1506)       \\
it               & 20330.9(38813) & 1311.1(3224)     & 15.5$\times$   & \textbf{1294.8}(3224)      \\
\bottomrule
\end{tabular}
\end{table}

\subsubsection{Parallel Index2core-based Algorithm Processing Time}

As shown in Table~\ref{tab:nbrcnthisto}, using the cnt value to locate the frontiers, \emph{CntCore} achieves an average speedup of 1.8$\times$ compared to \emph{NbrCore}. Additionally, by using the up-to-date \emph{histo} array to minimize neighbor access, \emph{HistoCore} achieves an average speedup of 8$\times$ compared to \emph{CntCore}. These results demonstrate the parallel potential of the \emph{Index2Core} paradigm, particularly with the use of \emph{HistoCore}.
In summary, the experiment demonstrates that \emph{HistoCore} can achieve the best performance in the \emph{Index2core}-based algorithms. 
 
\begin{table}[t]
\caption{The performance of \emph{NbrCore}, \emph{CntCore} and \emph{HistoCore}.}
\label{tab:nbrcnthisto}
\setlength{\tabcolsep}{8pt}
\centering
\small
\begin{tabular}{cccccc}
\toprule
dataset & NbrCore & CntCore & \textbf{HistoCore}  & Speed Up    & $l_2$   \\
\midrule
gow     & 57.6    & 28.5    & \textbf{3.1}       & 9.2$\times$  & 40   \\
ama     & 26.2    & 17.2    & \textbf{3.0}       & 5.7$\times$  & 78   \\
talk    & 323.5   & 139.0   & \textbf{14.0}      & 9.9$\times$  & 44   \\
goo     & 18.1    & 13.7    & \textbf{4.2}       & 3.2$\times$  & 24   \\
ber     & 640.0   & 361.8   &\textbf{ 31.0}     & 11.7$\times$ & 424  \\
ski     & 370.1   & 169.7   & \textbf{19.1}     & 8.9$\times$  & 64   \\
pat     & 84.1    & 98.4    & \textbf{16.2}     & 6.1$\times$ & 63   \\
in      & 573.1   & 849.7   & \textbf{40.9}      & 20.8$\times$ & 976  \\
dbl     & 48.1    & 59.7    & \textbf{17.8}     & 3.4$\times$  & 66   \\
woc     & 304.5   & 111.8   &\textbf{ 18.5}      & 6.0$\times$  & 164  \\
lj      & 502.3   & 344.9   & \textbf{115.2}     & 3.0$\times$  & 105  \\
wde     & 2601.7  & 896.1   & \textbf{219.6}     & 4.1$\times$  & 131  \\
hol     & 490.3   & 267.9   & \textbf{81.5}      & 3.3$\times$  & 59   \\
ork     & 2860.9  & 1686.0  & \textbf{567.3}     & 3.0$\times$  & 192  \\
tra     & 55480.3 & 14618.9 & \textbf{1425.6}    & 10.3$\times$ & 45   \\
ind     & 5485.1  & 5122.7  & \textbf{327.7}     & 15.6$\times$ & 1253 \\
uk      & 5697.0  & 3231.8  & \textbf{323.3}     & 10.0$\times$ & 588  \\
sina    & 7059.9  & 6098.4  & \textbf{788.0}     & 7.7$\times$  & 110  \\
twi     & 8348.7  & 5179.6  & \textbf{806.4}     & 6.4$\times$  & 84   \\
wien    & 9453.2  & 3191.1  & \textbf{886.9}     & 3.6$\times$  & 93   \\
ara     & 32193.1 & 15050.3 & \textbf{1226.2}    & 12.3$\times$ & 1739 \\
uk      & 27204.4 & 8446.9  & \textbf{1083.6}    & 7.8$\times$  & 351  \\
wb      & 43293.1 & 32613.0 & \textbf{4625.2}    & 7.1$\times$  & 2069 \\
it      & 68607.8 & 49933.2 & \textbf{4066.0}    & 12.3$\times$ & 3525 \\
\bottomrule
\end{tabular}
\end{table}


\subsubsection{Peel vs. Index2core}

The \emph{PeelOne} method with the dynamic frontier represents the optimal GPU performance of the \emph{Peel} paradigm, while \emph{HistoCore} represents the optimal GPU performance of the \emph{Index2Core} paradigm. We compare the performance of these two algorithms as shown in Table~\ref{tab: peelvsindex2core}. 
The result shows that 
\emph{PeelOne-dyn} outperforms \emph{HistoCore} by 2$\times$ speedup in 6 datasets and exhibits marginally better performance (less than 2$\times$ speedup) on 12 datasets. However, on the remaining 6 datasets, \emph{HistoCore} surpasses \emph{PeelOne-dyn}, achieving a performance enhancement of 1.1$\times$$\sim$3.2$\times$ speedup.
Through an in-depth study of these 6 datasets (with the fonts in bold in Table~\ref{tab: peelvsindex2core}), we found that the value of $l_2$ is significantly smaller than that of $l_1$. 
$l_1$ and $l_2$ represent the iteration number of the \emph{Peel}-based and \emph{Index2core}-based algorithms, respectively. 
This indicates that the primary performance advantage of \emph{HistoCore} lies in processing graphs with a fewer number of iterations.
The minimum value of iteration number for the \emph{Peel}-based algorithms is fixed (to the max coreness $k_{max}$ of a graph).
From the perspective of vertex convergence dependency, the \emph{Index2Core}-based algorithms compute the coreness in a top-down way so that the iteration may be smaller than the \emph{Peel}-based algorithms.
Furthermore, we can see that the value of $l_1$ ($k_{max}$) of the 6 datasets is relatively big, compared to graphs with a similar number of edges. 
Therefore, \emph{HistoCore} shows the performance advantages in processing the graphs with deeper hierarchical structures. In summary, the experiment demonstrates that the \emph{Index2core}-based algorithm \emph{HistoCore} and the \emph{Peel}-based algorithm \emph{PeelOne} have their own performance advantages on the datasets with different statistical properties. The parallel \emph{Index2Core} paradigm is also competitive with the parallel \emph{Peel} paradigm on GPU. 



\begin{table}[t]
\caption{The parallel \emph{Peel} paradigm vs. the \emph{Index2core} paradigm.}
\label{tab: peelvsindex2core}
\setlength{\tabcolsep}{8pt}
\centering
\small
\begin{tabular}{ccccc}
\toprule
dataset             & PO-dyn                & $l_1$                  & HistoCore              & $l_2$  \\
\midrule
loc-Gowalla      & 3.0                  & 51                  & 3.1                  & 40                  \\
amazon0601       & 1.0                  & 10                  & 3.0                  & 78                  \\
wiki-Talk        & {\ul \textbf{24.0}}  & {\ul \textbf{131}}  & {\ul \textbf{14.0}}  & {\ul \textbf{44}}   \\
web-Google       & 3.0                  & 44                  & 4.2                  & 24                  \\
web-BerkStan     & 14.8                 & 201                 & 31.0                 & 424                 \\
as-Skitter       & {\ul \textbf{22.9}}  & {\ul \textbf{111}}  & {\ul \textbf{19.1}}  & {\ul \textbf{64}}   \\
cit-Patents      & 10.0                 & 64                  & 16.2                 & 63                  \\
in-2004          & 22.0                 & 488                 &40.9                  & 976                 \\
dblp-author      & 7.0                  & 14                  & 17.8                 & 66                  \\
wikipedialink-oc & {\ul \textbf{59.8}}  & {\ul \textbf{1252}} & {\ul \textbf{18.5}}  & {\ul \textbf{164}}  \\
LiveJournal1     & 56.7                 & 372                 & 115.2                & 105                 \\
wikipedialink-de & 211.0                & 837                 & 219.6                & 131                 \\
hollywood-2009   & {\ul \textbf{136.7}} & {\ul \textbf{2208}} & {\ul \textbf{81.5}}  & {\ul \textbf{59}}            \\
com-Orkut        & 104.0                & 253                 & 567.3                & 192                 \\
trackers         & 1030.8               & 438                 & 1425.6               & 45                  \\
indochina-2004   & {\ul \textbf{514.7}} & {\ul \textbf{6869}} & {\ul \textbf{327.7}} & {\ul \textbf{1253}} \\
uk-2002          & 207.3                & 943                 & 323.3                & 588                 \\
soc-sinaweibo    & 467.6                & 193                 & 788.0                & 110                 \\
soc-twitter-2010 & {\ul \textbf{914.2}} & {\ul \textbf{1695}} & {\ul \textbf{806.4}} & {\ul \textbf{84}}   \\
wikipedialink-en & 690.1                & 1114                & 886.9                & 93                  \\
arabic-2005      & 869.2                & 3247                & 1226.2               & 1739                \\
uk-2005          & 437.7                & 588                 & 1083.6               & 351                 \\
webbase-2001     & 1387.2               & 1506                & 4625.2               & 2069                \\
it-2004          & 1294.8               &3224                 & 4066.0                & 3525               \\
\bottomrule
\end{tabular}
\end{table}

%% file: sec6_relatedworks.tex
\section{RELATED WORK}
\label{relatedworks}

\subsection{Core Decomposition In Different Settings}

\subsubsection{In-Memory Setting}
Seidman \textit{et al}.~\cite{peel-paradigm} first proposed the concept of $k$-core and the \emph{Peel} paradigm. In a subsequent advancement, Batagelj \textit{et al}.~\cite{BZ} propose the state-of-the-art serial \emph{Peel}-based algorithm (BZ) with a time complexity of $\mathcal{O}{(M)}$. 
Specifically, BZ employs three arrays: the $vertices$ array to store the vertices in ascending order of degree, the $bin$ array to store the starting position of each bin in the $vertices$ array, the $position$ array to store the position of each vertex in $vertices$ array. By iterating in ascending order of bin numbers, the algorithm removes the vertices belonging to the same bin, updates the degrees of neighbors, and maintains the order of the $vertices$ array by shifting the position of neighbors to smaller bins. 
The key contribution of the BZ algorithm is the reduction of the time complexity of the \emph{Peel} paradigm through bin sorting.

\subsubsection{Out Of Memory and Distributed Setting}
To process large-scale graphs that may not reside entirely in the main memory, Cheng \textit{et al}.~\cite{oom-core} introduce the \emph{Peel}-based EMcore, which consists of three key components: an efficient strategy for graph partitioning, an effective mechanism for estimating the upper bound of the core number of the vertices, and a recursive top-down core decomposition procedure. However, EMCore cannot bound the size of the memory for graph slices, meaning that it still loads most edges of the graph into main memory. In response, Wen \textit{et al}.~\cite{ dyna-core3} propose an \emph{Index2core}-based semi-external algorithm for core decomposition with guaranteed memory bound. When the size of the graph exceeds the capacity of a single machine, existing works~\cite{distributed-core, dyna-core6} concentrate on optimizing the \emph{Index2core} paradigm in a distributed setting, as it is well-suited for vertex-centric distributed computing frameworks.

\subsubsection{Multi-core Setting}
Two recent works are studying multi-core algorithms for core decomposition. Park~\cite{park-core} follows the \emph{Peel} paradigm and adopts a two-phase method in each sub-level. 
To avoid the synchronization overhead within the same core at the sub-level, PKC~\cite{pkc-core} assigns a local dynamic work-list queue called $buff$ to each thread. 

\subsubsection{GPU Setting} 
Zhang \textit{et al}.~\cite{GPU-core1} firstly implement \emph{Peel} and \emph{Index2core} on the GPU. They preliminarily study and compare the performance of \emph{General Parallel Peel} and the \emph{NbrCore} on GPU. The other work all focus on the \emph{Peel} paradigm on GPU. VETGA~\cite{GPU-core3} abstracts the \emph{General Parallel Peel} in terms of vector primitives, leveraging highly optimized GPU vector operations such as PyTorch and GraphBLAS~\cite{graphblas}. The GPU-hardwired work~\cite{GPU-core4} is currently the stat-of-the-art GPU implementation inspired by the fastest multicore implementatio, PKC. They significantly improve performance by utilizing block-level dynamic frontier queues. Furthermore, the classic Graph-based GPU system Gunrock~\cite{gunrock} has recently added the \emph{General Parallel Peel} algorithm.

\subsection{Core Decomposition in Different Graph Types}

Core decomposition has been extended to various types of graphs, including signed graphs~\cite{s-core2}, weighted graphs~\cite{w-core}, hypergraphs~\cite{h-core1, h-core2}, and others. In this paper, we focus on two highly regarded categories: directed graphs and uncertain graphs.

\subsubsection{Directed Graphs}
A $(k, l)$-core refers to a maximal subgraph in which each vertex has an in-degree and out-degree of at least $k$ and $l$, respectively. This concept serves as a significant graph model. Giatsidis \textit{et al}.~\cite{d-core2} are the first to extend the concept of the $k$-core to directed graphs. They introduce a \emph{Peel}-based D-core decomposition method for detecting and evaluating directed communities (subgraphs). Liao \textit{et al}.~\cite{d-core3} propose an \emph{Index2core}-based algorithm for D-core decomposition to handle large-scale directed graphs efficiently in distributed settings.

\subsubsection{Uncertain Graphs}

A $(k, \eta)$-core refers to a maximal subgraph where each vertex has a probability of at least $\eta$ to have a degree of at least $k$. Bonchi \textit{et al}.~\cite{uncertain-core1} are the first to study the efficient computation of the $k$-core on uncertain graphs and proposed a \emph{Peel}-based algorithm. Li \textit{et al}.~\cite{uncertain-core2} later propose an improved \emph{Peel}-based algorithm with a low time complexity. Yang \textit{et al}.~\cite{uncertain-core3} further improved the scalability of the $(k, \eta)$-core and introduce an \emph{Index2core}-based algorithm. Dai \textit{et al}.~\cite{uncertain-core4} revisit the previous research and address the inaccuracy issues with the recursive floating-point number division operations involved in the \emph{Peel}-based $(k, \eta)$-core algorithm.


\subsection{The Variants of Core Decomposition}

\subsubsection{Core Maintenance Problem}

The key problem is to maintain the coreness of each vertex and avoid recomputing the entire graph when new edges are inserted or removed. The basic idea is to determine the small induced subgraph based on the updated edges and recalculate the coreness of the vertices in this subgraph~\cite{dyna-core1}. Wen \textit{et al}.~\cite{dyna-core3} further propose an I/O efficient core maintenance algorithm based on the \emph{Index2core} paradigm. To explore the parallelism of core maintenance, Wang \textit{et al}.~\cite{dyna-core4, dyna-core5} propose an advanced edge set data structure for processing the updated edges in parallel. Weng \textit{et al}.~\cite{dyna-core6} investigate the scalability of core maintenance in distributed settings.

\subsubsection{Graph Structure Optimization and Personalized Query}
$k$-core is usually used as a network model to help analyze the robustness of the networks. To reinforce the networks, Core maximization aims to maximize the coreness gain of the whole or subgraph by anchoring a small number of vertices in the network~\cite{anchored-core0, anchored-core1, anchored-core2}. Correspondingly, Liu \textit{et al}.~\cite{anchored-core3} explore the vulnerability of the $k$-core structure by deleting edges and Zhou \textit{et al}.~\cite{anchored-core4} further explore this problem in network-attacking scenarios. 
To query and search the personalized $k$-core, Li \textit{et al}.~\cite{person-core0} aim to find the smallest $k$-core subgraph containing every query vertex. Chu \textit{et al}.~\cite{person-core1} study how to find the best $k$-core set according to a personalized community scoring metric.

%% file: sec7_conclusion.tex
\section{Conclusion and Future}
This paper proposes an efficient GPU acceleration framework, PICO, for all paradigms of $k$-core decomposition, including the optimal \emph{Peel}-based algorithm \emph{PeelOne} and the optimal \emph{Index2core}-based algorithm \emph{HistoCore}. The experiments show that \emph{HistoCore} can outperform the further optimized \emph{PeelOne}, which proves the great parallel potential of the \emph{Index2core} paradigm as same with the \emph{Peel} paradigm.

In the future, we aim to explore the hybrid core decomposition algorithm to achieve the best performance on all real-world networks. We will also extend our framework for real-world networks with rich semantics. Furthermore, to process super-big graphs, we intend to explore the performance of our framework in multi-GPU and out-of-GPU settings.

\label{sec:conclusion}